\documentstyle[twoside,psfig]{article}
\input{ijmpb-macros}

\begin{document}

\runninghead{Effect of Strong Correlation on the Electron-Phonon
 Interaction} 
{Effect of Strong Correlation on the Electron-Phonon
      Interaction}

\normalsize\textlineskip
\thispagestyle{empty}
\setcounter{page}{1}

\copyrightheading{}                     %{Vol. 0, No. 0 (1993) 000---000}

\vspace*{0.88truein}

\fpage{1}
\centerline{\bf EFFECT OF STRONG CORRELATION}
\vspace*{0.15truein}
\centerline{\bf ON THE ELECTRON-PHONON INTERACTION}
\vspace*{0.37truein}

\centerline{\footnotesize LILIA BOERI}
\vspace*{0.015truein}
\centerline{\footnotesize\it Dipartimento di Fisica, 
Universit\`a ``La Sapienza'', P.le Aldo Moro 2}
\baselineskip=10pt
\centerline{\footnotesize\it Roma, 00185, Italy} 
\baselineskip=10pt
\centerline{\footnotesize\it and INFM, Unit\`a Roma1}

\vspace*{10pt}
\centerline{\footnotesize EMMANUELE CAPPELLUTI}
\vspace*{0.015truein}
\centerline{\footnotesize\it Dipartimento di Fisica, 
Universit\`a ``La Sapienza'', P.le Aldo Moro 2}
\baselineskip=10pt
\centerline{\footnotesize\it Roma, 00185, Italy} 
\baselineskip=10pt
\centerline{\footnotesize\it and INFM, Unit\`a Roma1}

\vspace*{10pt}
\centerline{\footnotesize CLAUDIO GRIMALDI}
\vspace*{0.015truein}
\centerline{\footnotesize\it \'Ecole Polytechnique F\'ed\'erale, 
D\'epartment de microtechnique IPM}
\baselineskip=10pt
\centerline{\footnotesize\it Lausanne, CH-1015, Switzerland}

\vspace*{10pt}
\centerline{\normalsize and}

\vspace*{10pt}
\centerline{\footnotesize LUCIANO PIETRONERO}
\vspace*{0.015truein}
\centerline{\footnotesize\it Dipartimento di Fisica, 
Universit\`a ``La Sapienza'', P.le Aldo Moro 2}
\baselineskip=10pt
\centerline{\footnotesize\it Roma, 00185, Italy} 
\baselineskip=10pt
\centerline{\footnotesize\it and INFM, Unit\`a Roma1}
\vspace*{0.225truein}

\abstracts{High-$T_c$ superconductors are usually described as strongly 
correlated
electronic systems.
This feature deeply affects the one-particle and two-particle properties
of the system. In particular, a large incoherent background developes on the
top of a narrow quasi-particle peak in the one-electron spectral
function. We schematize this structure with a simple phenomenological
form. The corresponding Green's function is employed to calculate
the charge response of the system taking into account in a proper way
strong correlation effects. The effective charge interaction acquires
a structure in the exchanged momentum space with a predominance
of forward scattering, in agreement with previous numerical calculations.
The consequences of the momentum dependence of the interaction are discussed
in the framework of the nonadiabatic theory of superconductivity proposed for
the high-$T_c$ materials.}{}{}

\textlineskip                   %) USE THIS MEASUREMENT WHEN THERE IS
\vspace*{12pt}                  %) NO SECTION HEADING

\textheight=7.8truein
\setcounter{footnote}{0}
\renewcommand{\thefootnote}{\alph{footnote}}

\section{Introduction}

High Temperature Superconductors (HTSC) show a highly complex phenomenology, as 
it is evident from the extremely rich phase diagram; this issue, however, 
is not 
surprising since
in this class of materials the electronic bands are very narrow.  Kinetic 
energies are accordingly quite small, so that the system is sensitive to any 
kind of instability which may give rise to visible effects. 
In this situation,
it is clear that, in order to provide an effective description
of the properties of HTSC, electronic correlation may not be neglected.  

Strongly correlated systems show some common characteristic features.
On one hand the  spectral weight  associated with the coherent part
of the one-particle  spectral function is reduced with respect to
the uncorrelated case and meanwhile a broad
incoherent background arises. 
In addiction, the electronic dispersion of the
coherent peak is narrowed by correlation effects.
As a result, in a correlated system
electronic kinetic energies, characterized 
by the Fermi energies $E_F$,
may become comparable with phonon frequencies $\omega_{ph}$, leading to the
breakdown of the adiabatic hypothesis (i.e. $\omega_{ph}/ E_F \ll 1$)
on which conventional superconductivity theory rests. 

In addiction to the modification of one-particle properties, electronic 
correlation affects as well
the two-particle ones of the system. 
In particular different analytical approaches point out to a predominance 
of the 
forward scattering (small {\bf q}) 
in the charge density response function.\cite{ze} This feature, 
together with the above 
mentioned bandwidth reduction, 
makes Migdal's theorem \cite{migdal} unappliable.

In this paper we investigate the effects of electronic correlation on the 
electron-phonon coupling. The modulation of the electron-phonon scattering
induced by correlation results to be
particularly important in the Cooper channel leading to
new feature of the superconductivity properties.

\section{The Model}

Systems of correlated electrons are usually described in terms of a Hubbard 
Hamiltonian,
which can be considered a good paradigmatic model\cite{Vol}:
\begin{equation}
\hat{H}=\sum_{i,j,\sigma}t_{i,j}\hat{c}_{i,\sigma}^\dagger
\hat{c}_{j,\sigma} +U \sum_i\hat{n}_{i,\uparrow}\hat{n}_{i,\downarrow}.
\end{equation}

The behaviour of the system can easily be determined in the two limit cases 
$(t/U) 
\rightarrow \infty $  
and $(t/U) \rightarrow 0$. In the first case the Hubbard Hamiltonian reduces 
to the free-electron one: the system is a 
perfect Fermi gas, with energy levels $\varepsilon_{\bf k}$ and Fermi energy 
$E_F$.
In the second case the most favorable configuration is that in which the 
electrons are all ``localized'', 
i.e. no hopping between sites is possible. 
In the intermediate region of parameters one expects a coexistence of the two 
behaviours.

In terms of the spectral function $A({\bf k},\omega)$
the situation can be described as follows: when $(t/U) = \infty$, 
$A({\bf k},\omega)$ 
is a 
delta-function centered at the energy $\varepsilon_{\bf k}$, fulfilling the sum 
rule 
 $\sum_{\bf k}\int d\omega A({\bf k},\omega) e^{i\omega 0^+} = N$,
where $N$ is the total number of particles.
 As $(t/U)$ decreases, the delta-function, 
which represents itinerant electrons, will `lose weight', i.e. the coherent 
spectral weight will be such that 
$\sum_{\bf k}\int d\omega A_{co}({\bf k},\omega) e^{i\omega 0^+} =NZ$, 
with $Z < 1$. The spectral weight which is lost in the coherent part forms an 
incoherent 
(i.e. independent of ${\bf k}$) 
background of states which represents localized electrons and is such that 
$\sum_{\bf k}\int d\omega A_{inc}({\bf k},\omega) 
e^{i\omega 0^+} =(1-Z)N$.\cite{Kotliar}
Even though Hubbard model cannot be solved exactly, there are many approximate 
methods 
of solution which can be used to determine the dependence of $Z$ 
on ``physical'' 
quantities, such as the doping 
$\delta $ or $U$. The above description
retains its validity regardless of the particular approximation chosen.  

In particular, without any loss of generality, the 
Green's function 
of the system can be written as 
$G=G_{co} + G_{inc} $. Traditional analytical approaches usually 
focus on either 
of the two components, according 
to which properties they want to underline: itinerant properties (mean field 
slave-bosons, Gutzwiller) or 
insulating ones (Hubbard I approximation). Interplay between the two components 
can be recovered at a higher order 
level of approximation.\cite{raimondi} 

In order to deal with this problems we have recently introduced 
an approximation based on a modified mean field solution within the slave 
boson technique, which allows us to treat both components
on the same footing. 

%%%% The formal derivation will be presented elsewhere; 
%%%% the main results are summarized below.

The Green's function can be expressed in a particularly simple form:
\begin{equation}
G({\bf k},\omega) = Z \tilde{G}({\bf k},\omega) + (1-Z) \sum_{\bf k} 
\tilde{G}({\bf k},\omega),
\label{eqz}
\end{equation}
where $\tilde{G}$ is the standard mean-field slave boson solution describing 
only the coherent part. 
It is clear that $Z$ gives a parameter to estimate the degree of 
correlation of 
the system
($Z=1$ uncorrelated, $Z=0$ maximally 
correlated).

The above result can be quite easily proved by splitting the 
generic Green's function 
in a local and non-local part:

\begin{equation}
G(i,j;t)= -i \langle T_t c_i(t)c_j^\dagger \rangle 
[1 - \delta_{i,j}] - i \langle T_t c_i(t)c_i^\dagger \rangle 
\delta_{i,j}
\end{equation}

In the limit of infinite $U$ the slave bosons technique 
decomposes the real electron into two operators:

\[c_{i,\sigma}=b_i^\dagger f_{i,\sigma}.
\]

The standard mean-field approximation 
[$b_i(t) \rightarrow b \equiv \langle b_i(t) \rangle$] 
is now applied only
to the non-local part of the Green's function, while in the 
local part the no-double occupancy constraint can be exactly implemented
by the relation

\[b_ib^\dagger_i f_i^\dagger (t)f_i=f_i^\dagger (t)f_i.
\]

After some simple algebra, we obtain therefore:

\begin{equation}
G(i,j;t)=-ib^2 
\langle T_t f_{i,\sigma}(t)f^\dagger_{j,\sigma}\rangle
-i[1 - b^2] \langle T_t f_{i,\sigma}(t)f^\dagger_{i,\sigma}\rangle \delta_{i,j},
\end{equation}

which, written in ${\bf k}$, gives eq. (\ref{eqz}) with $Z=b^2$.

It can be shown that the retarded Green's function 
of this form preserves the total 
spectral weight (
$\sum_{\bf k} \int d \omega \mbox{Im}G_R( {\bf k}, \omega) = N$). As a 
consequence Luttinger's theorem is
naturally fulfilled, contrary to what happens with other approximations. 
This function can then be used
as a basic ingredient to build up a diagramatic theory.

In this context we focus on the study of how charge-density response (namely 
electron-phonon interaction) is affected
by electronic correlation. 

\section{Electron-phonon interaction: predominance of forward scattering}

We wish to study how the effective electron-phonon coupling 
function $g^2({\bf q},\omega)$,  is affected by electronic correlation.
This issue has already been addressed with different analytical techniques 
focusing on charge fluctuations around 
mean-field solutions.\cite{ze} The results are in substantial agreement and show that 
electronic correlation induces 
a structure of the electron-phonon coupling which is strongly peaked around 
${\bf q}=0$ leading to a predominance 
of forward scattering. This structure can be thought of as resulting by the
poorer screening provided by the correlated electron system with respect to
a normal metal, where highly itinerant electrons effectively screen out
any charge modulation.

In this work we introduce an alternative and more intuitive way to describe 
correlation effects
in an electron-phonon system based on the use of the Green's function 
in eq. (\ref{eqz}).

Let us consider the screening of 
the bare electron-phonon interaction $g_0$ by the electron-electron scattering.
According the standard RPA approximation, the renormalized
$g^2$(q) can be written as:
\begin{equation}
g^2({\bf q},\omega) = 
\frac{g_0^2({\bf q},\omega)} {1-V({\bf q})\Pi ({\bf q},\omega)},
\end{equation}
where V({\bf q}) is the Coulomb repulsive potential between electrons and 
$\Pi({\bf q},\omega)$ is the charge density ``bubble'' of the 
correlated system:
\begin{equation}
\Pi({\bf q}, \omega) = \sum_{\bf k} \int_{-\infty}^{+\infty}   \left( \frac{d 
\omega'}{2 \pi} \right) 
 G({\bf k}, \omega') G({\bf k+q}, \omega + \omega').
\end{equation}

In a normal metal $\Pi(0)=-N(E_F)$, where $N(E_F)$ is the density of states
at the Fermi level, indicating that all the electrons at the Fermi surface
contribute to the screening. The total response can be express in term
of the ``Thomas-Fermi'' cut-off $k_{TF}$, which is usually larger
than the Brillouin zone, and the resulting screened electron-phonon interaction
is essentially ${\bf q}$-independent.

Things are different in a correlated system.
By using the expression of Green's function in eq. (\ref{eqz}), 
$\Pi({\bf q},\omega)$ can be rewritten in a coherent and incoherent part: 
\begin{equation}
\Pi({\bf q},\omega)=Z \Pi_{co}({\bf q},\omega)+\frac{(1-Z^2)}{Z} 
\Pi_{inc}(\omega),
\end{equation}
where, just as for $G$, $\Pi_{inc}(\omega)=
\sum_{\bf q}\Pi_{co}({\bf q},\omega)$.

The two components contribute in a different way to the screening
of the external charge. 
On a physical ground we expect that the part of the electrons
which exhibits itinerant behaviour 
will be screening the external charge
with a characteristic Thomas-Fermi cut-off
which scales with $Z$. On the other hand the localized states will
provide just a residual dynamical screening.

In this picture, the coherent part, which describes itinerant quasi-particles, 
can be described as a non-interacting renormalized system,  
by means of the standard 
Thomas-Fermi approximation: $\Pi_{co}({\bf q},\omega) = -N(E_F)$.
The factor $Z$ which multiplies $\Pi_{co}({\bf q},\omega)$ is due 
both to band renormalization 
and spectral weight reduction.
The incoherent part, which describes localized states, exhibits a more complex
behaviour as a function of the parameter $Z$ and of the exchanged frequency $\omega$.
When $\omega$ is comparable with the effective kinetic energies of the electrons $ZE_F$, namely
when the system is strongly correlated ($Z \rightarrow 0$) ,
the charge response of the localized states is  in counterphase,
yielding a negative screening.

\begin{figure}
\centerline{\psfig{figure=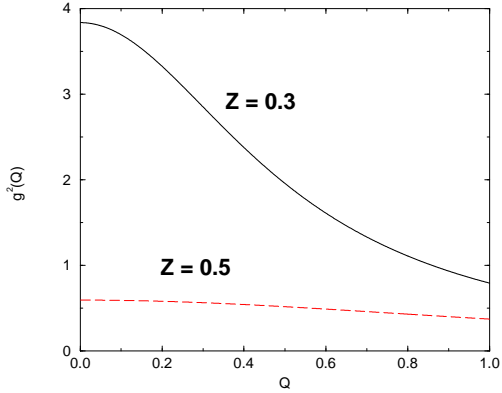,height=5.3 cm}}
\caption{Plot of the electron-phonon coupling-function vs $Q=(q/2k_F)$, for $(\omega_0/E_F)=0.3$.}
\label{peak}
\end{figure}

The resulting $g^2(q)$ can then be studied as a function of the degree of correlation of the system $Z$; 
in an intuitive picture $Z$ can be related to the hole-doping $\delta$ of the system through several approximations;
in the infinite-$U$ limit it can be shown that $Z \sim \delta$.
In Fig. \ref{peak} is plotted the renormalized electron-phonon interaction
as function of the exchanged momentum ${\bf q}$ for two different $Z$'s.
The Thomas-Fermi constant $k_{TF}$ has been chosen $k_{TF}=2k_F$.
As shown in the figure, $g^2(q)$ presents a sharp peak in the 
small-${\bf q}$ region,  
which corresponds to forward scattering.  A similar structure provides
a cut-off  $q_c$ for electron-phonon interactions 
in momentum-space which depends on the ``degree of correlation''
of the system.

\begin{figure}
\centerline{\psfig{figure=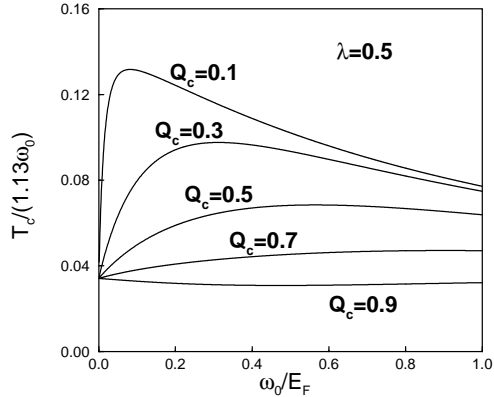 ,height=5.4 cm}}
\caption{Plot of $T_c$ calculated in the non-adiabatic theory of superconductivity, for various values of the momentum cut-off $Q_c=q_c/2k_F$.}
\label{temp}
\end{figure}

Predominance of forward scattering has important consequences
in the context of the electron-phonon theory of superconductivity.
In particular, it is clear that in this situation 
Migdal's theorem, which relies
on the small parameter $(\omega_{ph}/v_Fq)$, cannot be justified.\cite{migdal}
Vertex corrections, usually neglected in virtue of Migdals' theorem,
need therefore to be explicitely included.
In the past years, 
the extension of the theory of superconductivity
in the so-called ``nonadiabatic'' regime,
where first corrections
beyond Migdal's theorem are relevant, has been largely studied.\cite{pietro}
A major role is played by the momenta structure of the electron-phonon
interaction: for small ${\bf q}$'s the resulting
effects of the vertex corrections
is mainly positive, leading to an enhancement of the electron-phonon coupling
and to an increase of the superconducting critical temperature $T_c$ (see fig. \ref{temp}). 

\section{Conclusions}

In this work we present a rather simple and compact approach to the problem 
of electronic correlation in HTSC.
It is based on the observation that the main features of strongly correlated electron 
systems described by Hubbard-type models  are the reduction of the spectral weight 
associated with the itinerant part of the spectral function and the onset 
of an incoherent background of states.
Through the introduction of a phenomenological parameter $Z$ which describes
 the ``degree of correlation'' of the system we were able to treat the two parts of
 the spectral function on the same footing.
 We applied this model to the computation of the electron-phonon coupling function $g^2(q)$
 and found a structure which is strongly peaked in the small-${\bf q}$ region and 
 exhibits a strong dependence on $Z$ 
 
 The peaked structure of $g^2(q)$ provides an upper cut-off in momentum-space 
 for electron-phonon interactions, which
 leads to the breakdown of the adiabatic hypothesis in the momentum space $(\omega_{ph}/v_Fq)$. 
This points to the necessity of including electron-phonon vertex
corrections in Migdal-Eliashberg equations, as predicted by the non-adiabatic theory
 of superconductivity, which shows that $T_c$'s of HTSC are enhanced by vertex corrections when 
 the momentum ${\bf q}$ exchanged in electron-phonon interactions is small.


\begin{thebibliography}{99}


\bibitem{ze}
R. Zeyher and M. Kuli\'c, Phys. Rev. B {\bf 53} 2850 (1996);

M. Grilli and C. Castellani,Phys. Rev. B {\bf 50}, 16880 (1994).

\bibitem{migdal}
A. B. Migdal, Sov. Phys. JETP {\bf 7}, 996 (1958).

\bibitem{Vol}
D.Vollhardt, in {\em Correlated Electron Systems}, ed. V.J. Emery (World
Scientific, Singapore, 1993)

\bibitem{Kotliar}
A. Georges, G. Kotliar, W.Krauth and M.J. Rozenberg, Rev. Mod. Phys. {\bf 68},
13 (1996);


\bibitem{raimondi} R. Raimondi and C. Castellani,
Phys. Rev. B {\bf 48}, 11453 (1993).

\bibitem{pietro}
C. Grimaldi, L. Pietronero and S. Str\"assler,
Phys. Rev. Lett. {\bf 75}, 1158 (1995);

L. Pietronero, S. Str\"assler  and C. Grimaldi,
Phys. Rev. B {\bf 52}, 10516 (1995);

C. Grimaldi, L. Pietronero and S. Str\"assler,
Phys. Rev. B {\bf 52}, 10530 (1995).

\end{thebibliography}
\end{document}